\documentclass[aps,prl,floatfix,showpacs,twocolumn]{revtex4-1}
\usepackage{dcolumn}
\usepackage{bm}
\usepackage{color}
\usepackage{graphicx}
\usepackage{epsf}
\usepackage{pstricks}
\begin{document}

\title{Neutral weak current two-body contributions in inclusive scattering from $^{12}$C}
\author{A.\ Lovato$^{\, {\rm a,b} }$, S.\ Gandolfi$^{\, {\rm c} }$, J.\ Carlson$^{\, {\rm c} }$, Steven\ C.\ Pieper$^{\, {\rm b} }$,
and R.\ Schiavilla$^{\, {\rm d,e} }$}
\affiliation{
$^{\,{\rm a}}$\mbox{Argonne Leadership Computing Facility, Argonne National Laboratory, Argonne, IL 60439}\\
$^{\,{\rm b}}$\mbox{Physics Division, Argonne National Laboratory, Argonne, IL 60439}\\
$^{\,{\rm c}}$\mbox{Theoretical Division, Los Alamos National Laboratory, Los Alamos, NM 87545}\\
$^{\,{\rm d}}$\mbox{Theory Center, Jefferson Lab, Newport News, VA 23606}\\
$^{\,{\rm e}}$\mbox{Department of Physics, Old Dominion University, Norfolk, VA 23529}
}
\date{\today}

\begin{abstract}
An {\it ab initio} calculation of the sum rules of the
neutral weak response functions in $^{12}$C is reported, based on a
realistic Hamiltonian, including two- and three-nucleon potentials, and
on realistic currents, consisting of one- and two-body terms. 
We find that the sum rules of the response functions associated with
the longitudinal and transverse components of the (space-like) neutral current are
largest and that a significant portion ($\simeq 30$\%) of the calculated
strength is due to two-body terms.  This fact may have implications for the
MiniBooNE and other neutrino quasi-elastic scattering data on nuclei.

\end{abstract} 


\index{}\maketitle
In recent years, there has been a surge of interest in inclusive neutrino
scattering off nuclear targets, mostly driven by the anomaly observed
in the MiniBooNE quasi-elastic charge-changing scattering data on
$^{12}$C~\cite{Aguilar:2008}, i.e., the excess, at relatively low energy, of
measured cross section relative to theoretical calculations.  Analyses
based on these calculations have led to speculations that our present
understanding of the nuclear response to charge-changing weak probes
may be incomplete~\cite{Benhar:2010}, and, in particular, that the momentum-transfer 
dependence of the axial form factor of the nucleon may be rather different
from that obtained from analyses of pion electroproduction data~\cite{Amaldi:1979}
and measurements of neutrino and anti-neutrino reactions on protons and 
deuterons~\cite{Baker:1981,Miller:1982,Kitagaki:1983,Ahrens:1987}.

The accurate calculation of the weak inclusive response of a nucleus like $^{12}$C
is a challenging quantum many-body problem.  Its difficulty is compounded by the
fact that the energy of the incoming neutrinos is not known (in contrast, for example,
to inclusive $(e,e^\prime)$ scattering where the initial and final electron energies are
precisely known).  The observed cross section for a given energy and angle of the
final lepton results from a folding with the energy distribution of the incoming neutrino
flux and, consequently, may include contributions
from energy- and momentum-transfer regions of the nuclear response where different
mechanisms are at play: the threshold region, where the structure of the low-lying
energy spectrum and collective effects are important; the quasi-elastic region, which
is (naively, see below) expected to be dominated by scattering off individual nucleons;
and the $\Delta$ resonance region, where one or more pions are produced in the final
state.

In recent years, a number of studies have attempted to provide a description
of the nuclear weak response in this wide range of energy and momentum transfers.
They typically rely on a relativistic Fermi gas~\cite{Nieves:2004,Nieves:2004e} or relativistic mean
field~\cite{Caballero:2006,Caballero:2007} picture of the nucleus.  
Some, notably those of Ref.~\cite{Martini:2009,Martini:2010},
include correlation effects in the random-phase approximation induced by effective
particle-hole interactions in the $N$-$N$, $\Delta$-$N$, $N$-$\Delta$ and
$\Delta$-$\Delta$ sectors, use various inputs from pion-nucleus phenomenology, and
lead to predictions for electromagnetic and strong spin-isospin response
functions of nuclei, as measured, respectively, in inclusive $(e,e^\prime)$ scattering
and in pion and charge-exchange reactions, in reasonable agreement with data.

In the present manuscript, we report on a study of the neutral weak
response of $^{12}$C, based on a dynamical framework
in which nucleons interact among themselves with two- and three-body
forces and with external electroweak probes via one- and two-body
currents---elsewhere~\cite{Lovato:2013}, we have referred to this framework
as the standard nuclear physics approach (SNPA).  While SNPA allows for
an {\it ab initio} treatment of  the nuclear response in the threshold and
quasi-elastic regions and, as such, constitutes a significant improvement
over the far more phenomenological approaches mentioned above, it has
nevertheless severe limitations: it cannot describe---at least, in its present
formulation---the $\Delta$-excitation peak region, since no mechanisms
for (real) single- and multi-pion production are included in it.  However, the above
proviso notwithstanding, the sum rules of weak neutral response functions, which
we consider here, should provide useful insights into the nature of the strength
seen in the quasi-elastic region and, in particular, into the role of two-body
terms in the electroweak current.

The differential cross section for neutrino ($\nu$) and antineutrino ($\overline{\nu}$)
inclusive scattering off a nucleus---the processes $A(\nu_l,\nu_l^\prime)$ and
$A(\overline{\nu}_l,\overline{\nu}_l^\prime)$ induced by the neutral weak current
(NC)---can be  expressed in terms of five response functions as follows~\cite{Shen:2012}
\begin{eqnarray}
\!\!\!&&\!\!\!\left(\frac{ {\rm d}\sigma}{ {\rm d}\epsilon^\prime {\rm d}\Omega}\right)_{\nu/\overline{\nu}}
= \frac{G_F^{\, 2}}{2\pi^2}\, k^\prime \epsilon^\prime \, {\rm cos}^2 \frac{\theta}{2}\Bigg[
R_{00} +\frac{\omega^2}{ q^2}\, R_{zz} -\frac{\omega}{q} R_{0z} \nonumber \\
\!\!\!&&\!\!\!+
\left( {\rm tan}^2\frac{\theta}{2}+\frac{Q^2}{2\, q^2}\right) R_{xx}
\mp {\rm tan}\frac{\theta}{2}\, \sqrt{  {\rm tan}^2\frac{\theta}{2}+\frac{Q^2}{ q^2} } \, R_{xy}
\Bigg] \ ,\nonumber
\label{eq:xsw}
\end{eqnarray}
where $G_F=1.1803\times 10^{-5}$ GeV$^{-2}$ is the Fermi constant~\cite{Towner:1999}
and the $-$ ($+$) sign in the last term applies to the $\nu$ ($\overline{\nu}$)
reaction.  The neutrino initial and final four-momenta are $k^\mu=(\epsilon, {\bf k})$
and $k^{\mu \,\prime}=(\epsilon^\prime,{\bf k}^\prime)$, and its energy and momentum
transfers are defined as $\omega=\epsilon-\epsilon^\prime$ and ${\bf q}={\bf k}-{\bf k}^\prime$.
The scattering angle and four-momentum transfer are denoted by $\theta$
and $Q^2$, respectively, with $Q^2=q^2-\omega^2 > 0$.  The nuclear response functions
are schematically given by (explicit expressions are listed in Eqs.~(2.5)--(2.9) of Ref.~\cite{Shen:2012})
\begin{eqnarray}
R_{\alpha\beta}(q,\omega) &\sim & \overline{\sum_{i}}  \sum_f \delta( \omega\!+\!m_A\!-\!E_f) 
\langle f \! \mid j^\alpha({\bf q},\omega) \mid \! i \rangle \nonumber\\
&&\times \langle f\! \mid j^\beta({\bf q},\omega) \mid \! i \rangle^* \ ,
\label{eq:r5}\nonumber
\end{eqnarray}
where $\mid \!  i \rangle$ and $\mid \!\! f\rangle$ represent the initial
ground state and final scattering state of the nucleus of energies $m_A$ and 
$E_f= \sqrt{q^2+m_f^2}$; here, $m_A$ and $m_f$ denote, respectively, the rest mass
and internal excitation energy (including the masses of the constituent nucleons).
The three-momentum transfer ${\bf q}$ is taken along the $z$-axis (i.e., the
spin-quantization axis), and $j^\mu({\bf q},\omega)$ is the NC time component
for $\mu=0$ or space component for $\mu=x,y,z$.  Lastly, an
average over the initial nuclear spin projections is implied.

The NC is given by 
\begin{eqnarray}
j^\mu =-2\, {\rm sin}^2\theta_W\, j^\mu_{\gamma, S} + (1-2\, {\rm sin}^2\theta_W) \, j^\mu_{\gamma, V} 
+\, j^{\mu5}_V \ , \nonumber
\end{eqnarray}
where $\theta_W$ is the Weinberg angle (${\rm sin}^2\theta_W=0.2312$~\cite{PDG}), $j^\mu_{\gamma,S}$
and $j^\mu_{\gamma,V}$ denote, respectively,
the isoscalar and isovector components of the electromagnetic current, and $j^{\mu5}_{V}$ denotes the isovector
component of the axial current.  Isoscalar contributions to $j^\mu$ associated with strange quarks are ignored,
since experiments at Bates~\cite{Spayde:2000,Spayde:2004,Spayde:2005} and JLab~\cite{Ahmed:2012,Aniol:2004,Acha:2007} have found them to be very small.

Explicit expressions for the nuclear electromagnetic current $j^\mu_\gamma$ are reported
in Ref.~\cite{Shen:2012} and were used in our recent study of the
charge form factor and longitudinal and transverse sum rules of electromagnetic response
functions in $^{12}$C~\cite{Lovato:2013}.  In the SNPA they
lead to a satisfactory description of a variety of electro- and photo-nuclear observables
in systems with $A \leq 12$, ranging from static properties (charge radii, quadrupole moments,
and M1 transition widths) to charge and magnetic form factors to low-energy radiative capture
cross sections and to inclusive $(e,e^\prime)$ scattering in quasielastic kinematics at intermediate
energies~\cite{Pastore:2013,Marcucci:2008,Lovato:2013,Carlson:1998,Marcucci:2005,Carlson:2002}.

A realistic model for the axial weak current $j^{\mu 5}_V$ includes one- and two-body
terms (see Ref.~\cite{Shen:2012} for a recent overview).  The former follow from a non-relativistic
expansion of the single-nucleon four-current, in which corrections proportional to $1/m^2$
($m$ is the nucleon mass) are retained.  The time component of the two-body axial current
includes the pion-exchange term whose structure and strength are determined by soft-pion
theorem and current algebra arguments~\cite{Kubodera:1978}.  Its space components consist
of contributions associated with $\pi$- and $\rho$-meson exchanges, the axial $\rho\pi$ transition
mechanism, and a $\Delta$ excitation term (treated in the static limit).  The values for the $\pi$-
and $\rho$-meson coupling constants are taken from the CD-Bonn one-boson-exchange
potential~\cite{Machleidt:2001}.  Two different sets of cutoff masses $\Lambda_\pi$ and $\Lambda_\rho$
are used to regularize the $r$-space representation of these operators~\cite{Shen:2012}: in the first set
(Set I) the $\Lambda_\pi$ and $\Lambda_\rho$ values ($\Lambda_\pi$=$\Lambda_\rho$=1.2 GeV)
are in line with those extracted from the effective $\pi$-like and $\rho$-like exchanges implicit
in the Argonne $v_{18}$ (AV18) two-nucleon potential~\cite{Wiringa:1995}, while in the second
set (Set II) they are taken from the CD-Bonn potential ($\Lambda_\pi$=1.72 GeV and
$\Lambda_\rho$=1.31 GeV).  In the $N$ to $\Delta$ current, the value for the transition
axial coupling constant $(g_A^*)$ is determined by fitting the Gamow-Teller matrix element
of tritium $\beta$-decay in a calculation~\cite{Marcucci:2012,Marcucci:2011}  based on $^3$H/$^3$He wave
functions corresponding to the AV18 and Urbana IX (UIX) three-nucleon~\cite{Pudliner:1995}
potentials and on the present model for the axial current ($g^*_A$=0.614 $g_A$ with Set I
and $g^*_A$=0.371 $g_A$ with Set II).

The $\omega$-dependence in the current $j^\mu$ enters through the dependence on $Q^2$
of the electroweak form factors of the nucleon and $N$-to-$\Delta$ transition. 
We fix $\omega$ at the quasielastic peak energy, $\omega_{\rm qe}=\sqrt{q^2+m^2}-m$,
and evaluate these form factors at $Q^2_{\rm qe}=q^2-\omega_{\rm qe}^2$. Sum rules
of NC response functions, defined as
\[
S_{\alpha\beta} (q)\!=\! C_{\alpha\beta} \int_{\omega_{\rm el}}^\infty 
{\rm d}\omega\, R_{\alpha\beta}(q,\omega) \ ,
\]
can then be expressed as ground-state expectation values of the type
\begin{eqnarray}
S_{\alpha\beta}(q)\!\!&=&\!\!C_{\alpha\beta} \overline{\sum_i}\,
 \langle i | j^{\alpha \dagger}({\bf q}) j^\beta({\bf q})\!
 +\!(1\!-\!\delta_{\alpha\beta})\,  j^{\beta \dagger}({\bf q}) j^\alpha({\bf q}) |i\rangle \nonumber \\
S_{xy}(q)\!\!&=&\!\!C_{xy} \overline{\sum_i}\, {\rm Im}\, 
 \langle i | j^{x \dagger}({\bf q}) j^y({\bf q})
 -  j^{y \dagger}({\bf q}) j^x({\bf q}) |i\rangle \nonumber 
\end{eqnarray}
where $\omega_{\rm el}=\sqrt{q^2+m_A^2}-m_A$ is the energy transfer
corresponding to elastic scattering, the $C_{\alpha\beta}$'s are convenient
normalization factors (see below), $\alpha \beta=00$, $zz$, $0z$, and $xx$, and
for $\alpha\beta=xx$ the expectation value of
$j^{x \dagger} j^x+j^{y \dagger} j^y$ is computed.  Note that the sum rules
as defined above include the elastic and inelastic contributions; the former
are proportional to the square of electroweak form factors of the nucleus.
In the large $q$ limit, these nuclear form factors decrease rapidly with $q$, and the
sum rules reduce to the incoherent sum of single-nucleon contributions.  The
normalization factors $C_{\alpha\beta}$ are chosen such that
$S_{\alpha\beta}(q\rightarrow \infty) \simeq 1$, for example
\[
C^{-1}_{xy}=-\frac{q}{m}\, G_A(Q^2_{\rm qe})\left[ Z\, 
\widetilde{G}_M^p(Q^2_{\rm qe})-N\, \widetilde{G}_M^n(Q^2_{\rm qe})\right] \ ,
\]
where $Z$ ($N$) is the proton (neutron) number, $G_A$ is the weak axial form factor
of the nucleon normalized as $G_A(0)=g_A$ ($g_A$=1.2694~\cite{PDG}), and
$\widetilde{G}_M^p=\left(1-4\, {\rm sin}^2 \theta_W\right)
G_M^p/2-G_M^n/2$ and $\widetilde{G}_M^n=\left(1-4\, {\rm sin}^2 \theta_W\right)
G_M^n/2-G_M^p/2$ are its weak vector form factors (here, $G_M^p$ and $G_M^n$
are the ordinary proton and neutron magnetic form factors, determined from fits
to elastic electron scattering data off the proton and deuteron and normalized
to the proton and neutron magnetic moments:
$G_M^p(0)=\mu_p$ and $G_M^n(0)=\mu_n$).
\begin{figure}[bth]
\includegraphics[width=8cm]{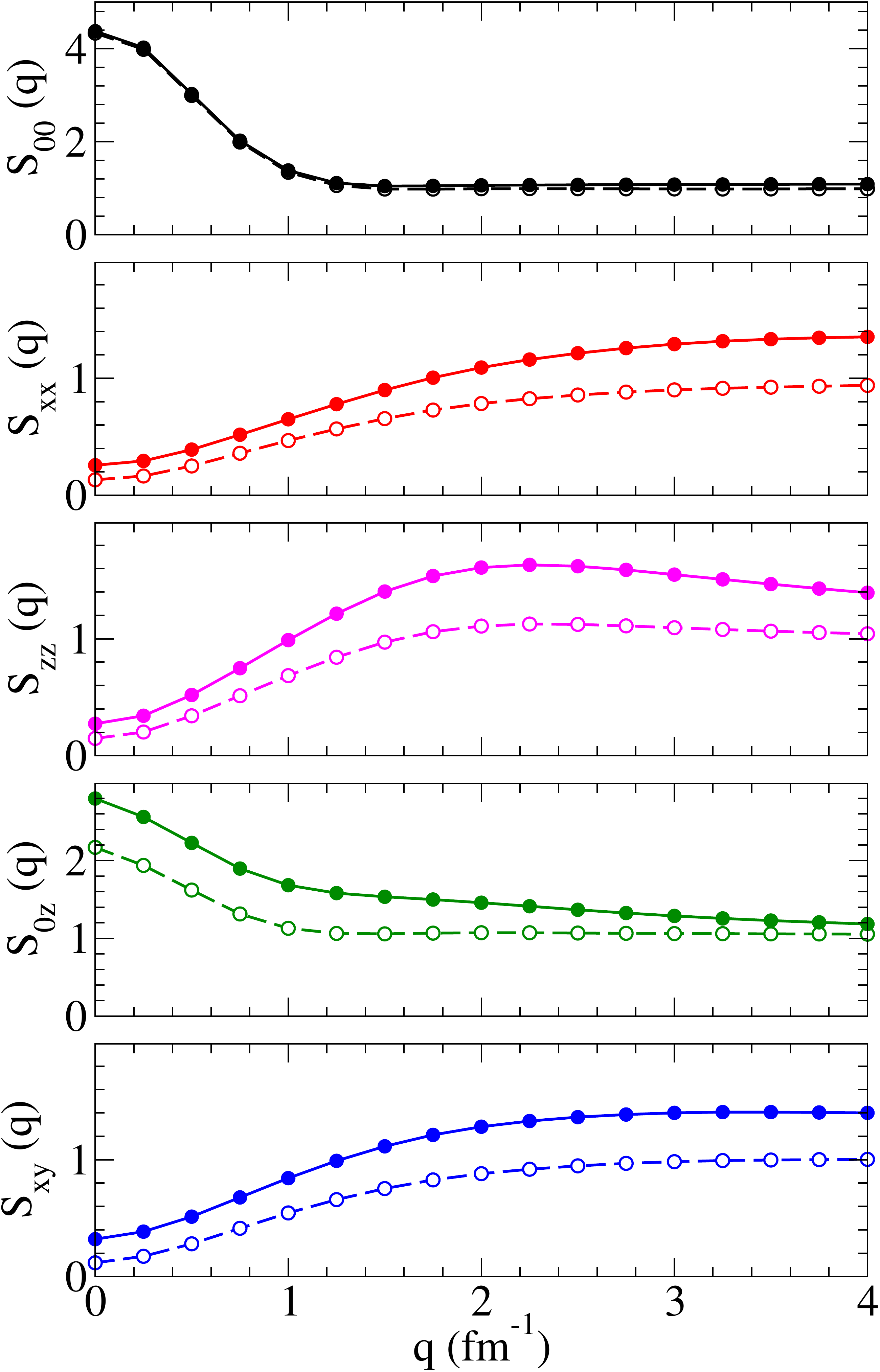}
\caption{(Color online) The sum rules $S_{\alpha\beta}$ in $^{12}$C,
corresponding to the AV18/IL7 Hamiltonian and obtained with one-body
only (dashed lines) and one- and two-body (solid lines) terms in the NC.}
\label{fig:f2}
\end{figure}
\begin{figure}[bth]
\includegraphics[width=8cm]{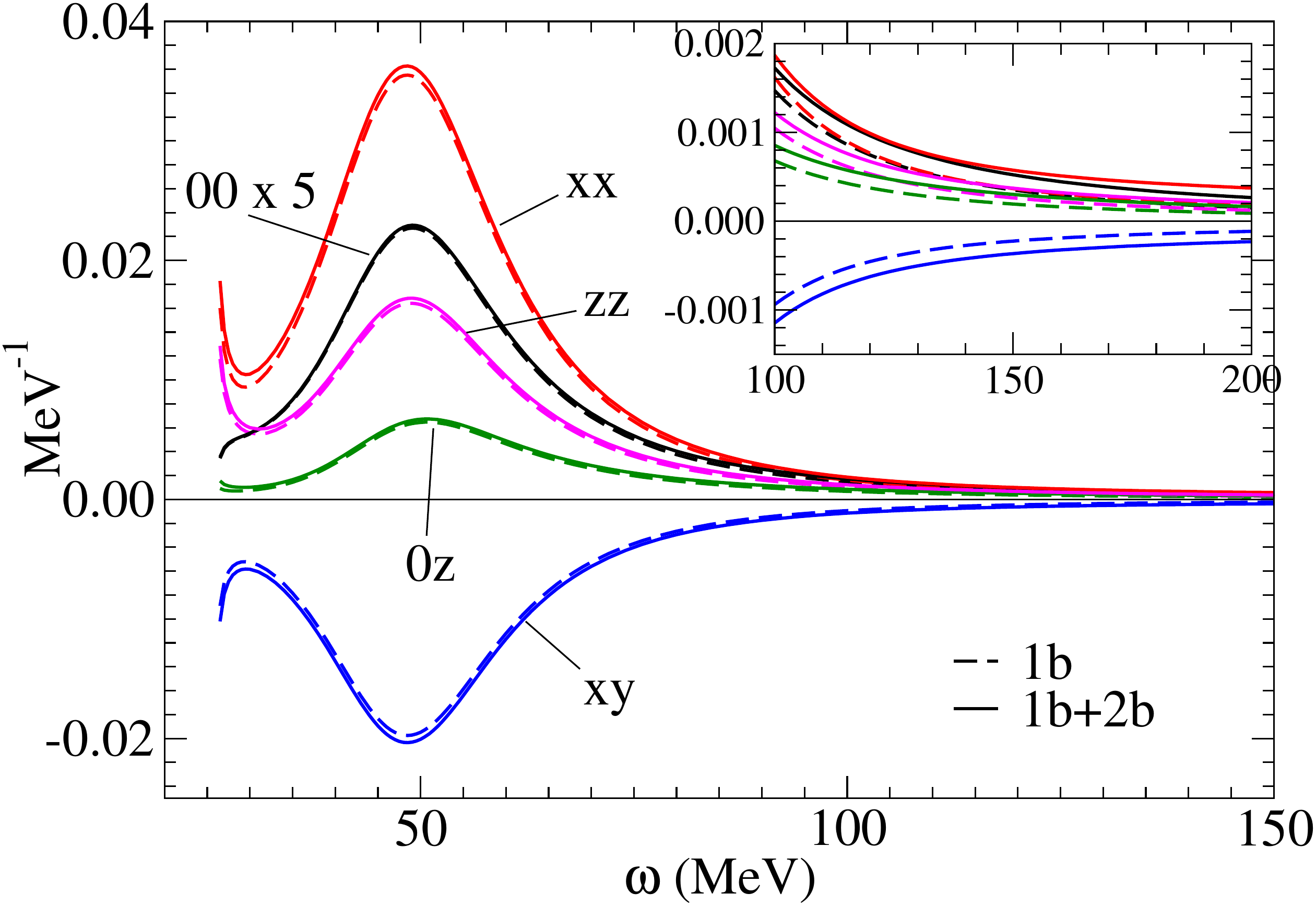}
\caption{(Color online) The response functions $R_{\alpha\beta}$ in the
deuteron at $q=300$ MeV/c computed using AV18 and obtained
with one-body only (dashed lines) and one- and two-body (solid lines)
terms in the NC.  The inset shows the tails of $R_{\alpha\beta}$ in the $\omega$-region
well beyond the quasi-elastic peak.}
\label{fig:f1}
\end{figure}

The ground-state wave function of $^{12}$C is obtained from a Green's
function  Monte Carlo (GFMC) solution of the Schr\"odinger equation including
the Argonne $v_{18}$ (AV18) two-nucleon~\cite{Wiringa:1995} and Illinois-7 (IL7)
three-nucleon~\cite{Pieper:2008} potentials.  The wave function is evolved in imaginary
time via a GFMC propagation starting from a variational wave function that contains
both explicit $\alpha$-clustering and the five possible $J^\pi$=0$^+$ p-shell
states.  The predicted ground-state energy, rms charge radius, and charge
form factor have been found to be in excellent agreement with experimental
data~\cite{Lovato:2013}.

The sum rules $S_{\alpha\beta}(q)$ in $^{12}$C are shown in Fig.~\ref{fig:f2}:
results $S^{\rm 1b}$  ($S^{\rm 2b}$) corresponding to one-body (one- and two-body)
terms in the NC are indicated by the dashed (solid) lines.  The two-body axial currents are
those of Set I; we find that Set II leads to very
similar results.  Note that both $S_{\alpha\beta}^{\rm 1b}$ and $S_{\alpha\beta}^{\rm 2b}$
are normalized by the (same) factor $C_{\alpha\beta}$, which makes
$S^{\rm 1b}_{\alpha\beta}(q) \rightarrow 1$ in the large $q$ limit.  In the
small $q$ limit, $S^{\rm 1b}_{00}(q)$ and $S^{\rm 1b}_{0z}(q)$ are much
larger than $S^{\rm 1b}_{\alpha\beta}$ for $\alpha\beta \neq 00,0z$.  In a simple
$\alpha$-cluster picture of $^{12}$C, one would expect
$S^{\rm 1b}_{\alpha\beta}(^{12}{\rm C})/C_{\alpha\beta}(^{12}{\rm C})
 \simeq 3\,S^{\rm 1b}_{\alpha\beta}(^{4}{\rm He})/C_{\alpha\beta}(^4{\rm He})$,
as is indeed verified in the actual numerical calculations to within a few \%, except for
$S^{\rm 1b}_{00}/C_{00}$ and $S_{0z}^{\rm 1b}/C_{0z}$ at low
$q  \lesssim 1$ fm $^{-1}$, where these quantities
are dominated by the elastic contribution
scaling as $A^2$. 
In the $\alpha$ particle, the operators $j^{0\, \dagger} j^0$ and $(j^{0\,\dagger} j^z+j^{z\,\dagger} j^0)$
can connect its dominant S-state components in the left and right
wave functions, while the remaining operator combinations cannot and
only contribute through S-to-D, D-to-S, and D-to-D transitions---D is the
D-state component, which has a probability of $\simeq 15\%$.

Except for $S_{00}^{\rm 2b} (q)$, the $S_{\alpha\beta}^{\rm 2b} (q)$ sum rules are considerably
larger than the $S_{\alpha\beta}^{\rm 1b} (q)$, by as much as 30-40\%. This 
enhancement was not seen in calculations of neutrino-deuteron scattering~\cite{Shen:2012};
the deuteron $R_{\alpha\beta} (q, \omega)$ 
response functions at $q = 300$ MeV/c are displayed in Fig. 2 
(note that $R_{00}$ is multiplied by a factor of 5). 
Two-body current contributions in the deuteron amount to only a few percent  at the 
top of the quasielastic peak of the 
(largest in magnitude) $R_{xx}$ and $R_{xy}$, but become increasingly 
more important in the tail of these response functions, 
consistent with the notion that this region is dominated by 
two-nucleon physics~\cite{Lovato:2013}. The very weak binding of the 
deuteron dramatically reduces the impact of two-nucleon currents, 
which are important only when two nucleons are within 1--2 inverse pion masses.

Correlations in $np$ pairs in nuclei with mass number A$\geq$3 are
stronger than in the deuteron. The two-nucleon density 
distributions in deuteron-like ($T$=0 and $S$=1) pairs are proportional 
to those in the deuteron for separations up to $\simeq$ 2 fm, and this 
proportionality constant, denoted as $R_{Ad}$ in Ref.~\cite{Forest:1996}, is
larger than $A/2$ (in $^4$He and $^{16}$O the calculated values of $R_{Ad}$ 
are 4.7 and 18.8, respectively).  Similarly, 
experiments at BNL~\cite{Piasetzky:2006}  and JLab~\cite{Subedi:2008} 
find that exclusive measurements of back-to-back pairs in 
$^{12}$C at relative momenta around $2\ {\rm fm}^{-1}$ are strongly 
dominated by $np$ (versus $nn$ or $pp$) pairs.  In this range and in the
back-to-back configuration,
the relative-momentum distribution of $np$ pairs is an order of magnitude
larger than that of $pp$ (or $nn$) pairs because of tensor correlations
induced by pion exchange.  The tensor force plays a larger role
in $np$ pairs where it can act in relative S-waves, while it acts only in
relative P-waves (and higher partial waves) in $nn$ and $pp$ pairs~\cite{Schiavilla:2007,Wiringa:2008}. 
We find that the enhancement in the weak
response due to two-nucleon currents is dominated
by $T$=0 pairs, much as was found previously 
in the case of the electromagnetic transverse response~\cite{Carlson:2002}.
For $S_{xx}$ and $S_{xy}$, the enhancement from $T$=1 $np$ pairs becomes
appreciable for $q\gtrsim 1\ {\rm fm}^{-1}$, while still remaining below $\simeq 15\%$ of that
due to $T$=0 pairs.  For $S_{zz}$, contributions from $T$=1 $np$ pairs are
larger at $q \simeq 1\ {\rm fm}^{-1}$, where they are about $30\%$ of those due to $T$=0
pairs. As for $S_{0z}$, at small momentum transfer the $T$=1 $np$-pair contributions  
are negative and interfere destructively with the $T$=0 ones. 

The increase due to two-nucleon currents is quite 
substantial even down to small momentum transfers. 
At $q \simeq 1\ {\rm fm}^{-1}$, the  enhancement is about 50\% relative to
the one-body values.  In general, the additional
contributions of the two-nucleon currents ($j_{\rm 2b}$)
to the sum rules are given by a combination of interference with 
one-body currents ($j_{\rm 1b}$), matrix elements of the type 
$\langle i \! \mid \! j_{\rm 1b}^\dagger\, j_{\rm 2b} \! \mid \! i \rangle 
+ \langle i\! \mid\!  j_{\rm 2b}^\dagger\, j_{\rm 1b}\! \mid\! i \rangle$,
and contributions of the type 
$\langle i \! \mid\! j_{\rm 2b}^\dagger\, j_{\rm 2b} \! \mid \! i \rangle$. 
At low momentum transfers we find the dominant contributions 
are of the latter $\langle i\!\mid j_{\rm 2b}^\dagger\, j_{\rm 2b} \mid\! i\rangle$ type, 
where the same pair is contributing in both left and right operators. 
One would expect the matrix element of any short-ranged two-body operator in 
$T,S = 0,1$  $np$ pairs, like the two-body weak currents under consideration here, 
to scale as $R_{Ad}$. Enhancements of the response due to two-nucleon currents 
could be important in astrophysical settings, where the neutrino energies typically 
range up to 50 MeV. A direct calculation of the $^{12}$C response 
functions is required to determine whether the strength of the 
response at low $q$ extends to the low energies kinematically 
accessible to astrophysical neutrinos.

At higher momentum transfers the interference between 
one- and two-nucleon currents plays a more important role.
The larger momentum transfer in the single-nucleon current connects the 
low-momentum components of the ground-state wave function directly with the 
high-momentum ones through the two-nucleon current.
For nearly the same Hamiltonian as is used here, there is a 10\% probability
that the nucleons have momenta greater than 2~fm$^{-1}$
implying that $\approx 30\%$ of the wave function amplitude 
is in these high-momentum components~\cite{Wiringa:2014}.  The contribution of
$np$ pairs remains dominant at high momentum transfers, and matrix
elements of the type $\langle i \!\mid [\, j_{\rm 1b} (l) 
+ j_{\rm 1b} (m) ]^\dagger j_{\rm 2b} (lm) \mid\! i \rangle + {\rm c.c.}$
at short distances between nucleons $l$ and $m$ are critical.
\begin{figure}[bth]
\includegraphics[width=8cm]{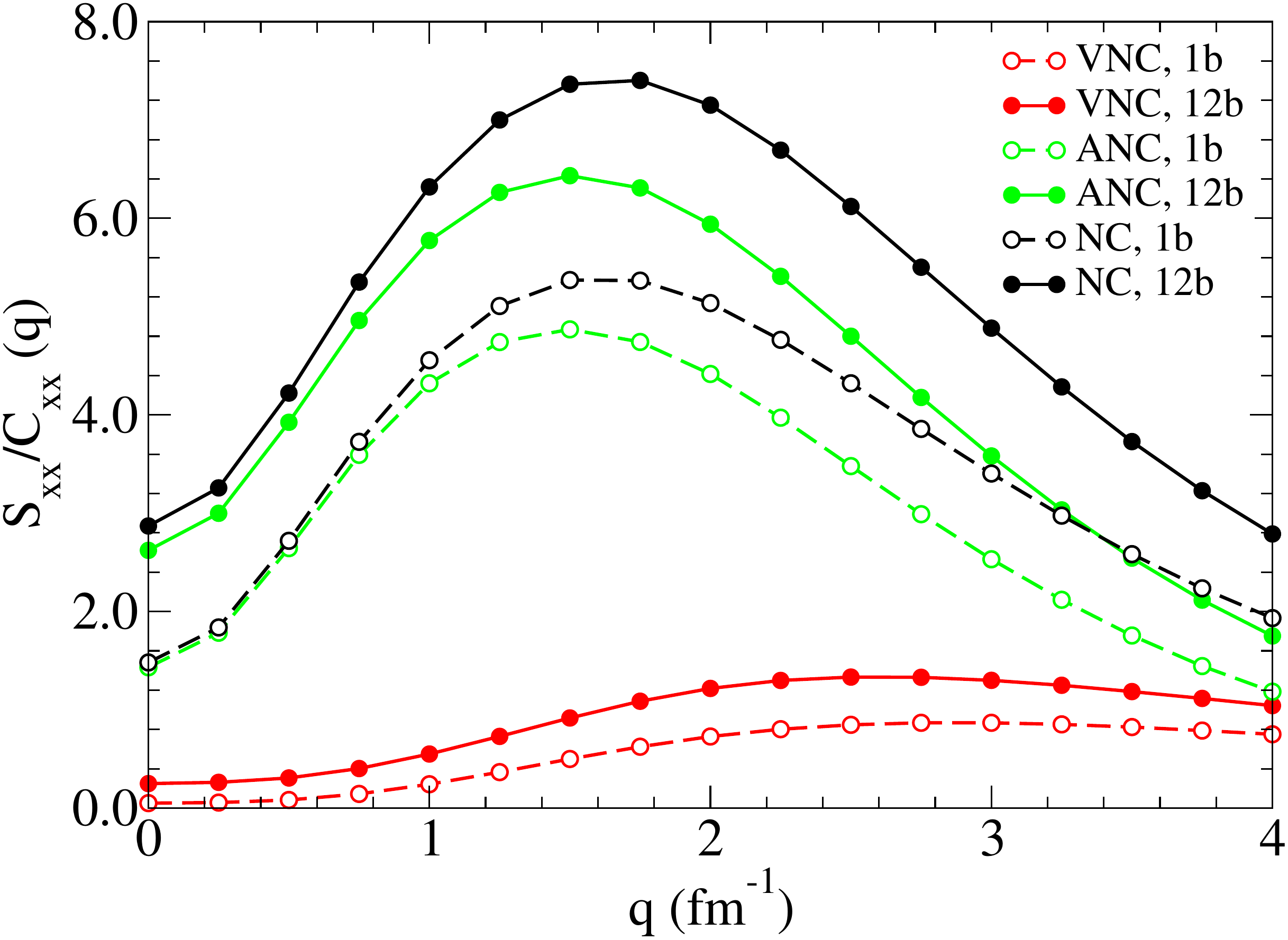}
\caption{(Color online) The $S_{xx}/C_{xx}$ sum rules obtained with the NC
(curves labeled NC) and either its vector (curves labeled VNC) or axial-vector (curves labeled ANC)
parts only. The corresponding one-body (one- and two-body) contributions are indicated
by dashed (solid) lines.  Note that the normalization factor $C_{xx}$ is not included.}
\label{fig:f3}
\end{figure}

In Fig.~\ref{fig:f3}, we show, separately, for the $S_{xx} / C_{xx}$
sum rule the contributions associated with the vector (VNC) and axial-vector (ANC) parts of the NC. 
We find that the ANC piece of the $S_{xx}$ sum rule has large two-body 
contributions (of the order of 30\% relative to the one-body). 
Similar results are found for the $0z$ and $zz$ sum rules; the $xy$ sum rule is nonzero because of 
interference between the VNC and ANC and vanishes in the limit in which only one or the other is considered.
The ANC two-body contributions in the sum rules studied here are much larger
than the contributions associated with axial two-body currents in weak charge-changing
transitions to specific states at low-momentum transfers, such as $\beta$-decays and
electron- and muon-capture processes involving nuclei with mass numbers
$A$=3--7~\cite{Marcucci:2011,Schiavilla:2002}, where they amount to a few \%
(but are nevertheless necessary to reproduce the empirical data).

In conclusion, the present study suggests that two-nucleon currents
generate a significant enhancement of the single-nucleon neutral weak current response,
even at quasi-elastic kinematics.  This enhancement is driven by
strongly correlated $np$ pairs in nuclei.  The presence of these
correlated pairs also leads to important interference effects
between the amplitudes associated with one- and two-nucleon
currents: the single-nucleon current can knock out two particles
from a correlated ground state, and the resulting amplitude
interferes with the amplitude induced by the action of the two-body
current on this correlated ground state.  The present results can be used
as constraints for more phenomenological approaches to the nuclear 
response, and to guide improvements to these models and experimental 
analyses of quasi-elastic scattering in neutrino experiments. 

\acknowledgments
  
Under an award of computer time provided by the INCITE program,
this research used resources of the Argonne Leadership Computing
Facility at Argonne National Laboratory, which is supported by the
Office of Science of the U.S. Department of Energy under contract
DE-AC02-06CH11357. 
We also used resources 
provided by Los Alamos Open Supercomputing, by the National Energy
Research Scientific Computing Center (NERSC), and by Argonne's LCRC.
This research is supported by the U.S.~Department of Energy, Office of
Nuclear Physics, under contracts DE-AC02-06CH11357 (A.L. and S.C.P.),
DE-AC02-05CH11231 (S.G.~and J.C.), DE-AC05-06OR23177 (R.S.), the
NUCLEI SciDAC program and by the LANL LDRD program.
\bibliography{biblio}

\end{document}